\documentclass[a4paper,12pt]{article}
\pdfoutput=1 
\usepackage{verbatim}

\usepackage{jheppub}

\usepackage{graphicx, subfigure}

\usepackage[T1]{fontenc}

\title{ Linear Quivers and $\mathcal{N}=1$ SCFTs from M5-branes}

\author{Ibrahima Bah$^{1}$ and Nikolay Bobev$^{2}$}
\affiliation{$^{1}$Institut de Physique Th\'{e}orique, CEA/ Saclay \\91191 Gif-sur-Yvette Cedex, France\\

$^{2}$Simons Center for Geometry and Physics\\
Stony Brook University \\
Stony Brook, NY 11794-3636, USA  \\

}

\emailAdd{Ibrahima.ba@cea.fr, nbobev@scgp.stonybrook.edu}

\abstract{We study a class of $\mathcal{N}=1$ quiver gauge theories build out of vector multiplets and matter multiplets in the fundamental and bifundamental representations. We argue that these theories flow to interacting SCFTs in the IR and calculate their central charges. We exhibit a type IIA brane construction which at low energies is described by these SCFTs. This also leads to a natural description of the theories in terms of M5-branes on a punctured sphere. }

\preprint{IPhT-T13/202}

\begin{document} 
\maketitle
\flushbottom

\bigskip
\bigskip
\bigskip
\bigskip

\section{Introduction}

Quiver gauge theories provide an interesting and very rich class of quantum field theories which arise naturally in string theory from branes placed at singularities or various brane intersections, see for example \cite{Douglas:1996sw,Hanany:1996ie, Giveon:1998sr}. 

In this paper we study a particular class of quiver gauge theories with $\mathcal{N}=1$ supersymmetry which are built out of $\mathcal{N}=1$ and $\mathcal{N}=2$ vector multiplets as well as ordinary matter multiplets. The quiver diagram encoding the field content of our theories has linear shape and hence we dub our theories \textit{linear quivers}. A key point in the construction is that we arrange the matter content of the theory and the superpotential such that we are left with precisely one non-anomalous $U(1)$ flavor symmetry in addition to the $U(1)_R$ R-symmetry. This is in the spirit of the field theory constructions in \cite{Bah:2011je,Bah:2011vv,Bah:2012dg} and  we will utilize many of the insights in these papers. We argue that the IR dynamics of the linear quivers is controlled by a set of interacting fixed points. A particular linear combination of the two global $U(1)$'s is then the superconformal R-symmetry in the IR. We find this linear combination using $a$-maximization \cite{Intriligator:2003jj}. This in turn facilitates the calculation of the central charges of the IR fixed points as well as the dimensions of some protected operators. 

Even though our discussion is inspired in part by the constructions in \cite{Bah:2011je,Bah:2011vv,Bah:2012dg}, we emphasize that we will not be using the strongly coupled isolated $T_N$ SCFT introduced in \cite{Gaiotto:2009we} as a building block for our quivers.  The $T_N$ itself can be defined by decoupling a set of $\mathcal{N}=2$ vector multiplets and hypermultiplets from linear quivers that preserve $\mathcal{N}=2$ supersymmetry \cite{Gaiotto:2009we,Gaiotto:2009gz}.  One of the motivations for studying the $\mathcal{N}=1$ linear quivers is to find possible $\mathcal{N}=1$ generalizations of the $T_N$ SCFT. We do not find such a generalization here but we believe that our construction is a useful step in this direction.

The construction of the linear quivers we study can be phrased entirely in the language of field theory without any reference to string theory or branes. However there are very natural type IIA constructions with D4- and NS5-branes which at low energies realize precisely the dynamics of our linear quivers. These brane constructions are in the spirit of \cite{Witten:1997sc} and are instrumental in understanding and interpreting the rules for building our linear quivers. Equipped with the type IIA picture we can follow the approach of \cite{Witten:1997sc} and \cite{Gaiotto:2009hg,Gaiotto:2009we} and take an M-theory limit. The linear quivers can then be thought of as an $\mathcal{N}=1$ twisted compactification of the $(2,0)$ theory on the world-volume of M5-branes on a punctured sphere. This limit allows also for a nice geometrization of many of the properties of the field theories of interest. Moreover it opens the way for an $\mathcal{N}=1$ generalization of the large class of $\mathcal{N}=2$ SCFTs constructed from M5-branes on Riemann surfaces \cite{Witten:1997sc, Gaiotto:2009hg,Gaiotto:2009we}. Many examples of such 4D $\mathcal{N}=1$ theories have already been discussed in the literature, see for instance \cite{Maruyoshi:2009uk,Benini:2009mz,Bah:2011je,Bah:2011vv,Bah:2012dg,Gadde:2013fma,Maruyoshi:2013hja}.  However we believe that the efforts so far only scratch the surface of a large structure underlying the space of 4D $\mathcal{N}=1$ SCFT's arising from M5-branes. 
%
%

The structure of this note is as follows. In the next section we present our setup and the rules for constructing linear quivers. In Section 3 we study their IR dynamics, argue that the theories flow to SCFTs and calculate the central charges and sueprconformal $R$-symmetry of the fixed points. The intersecting brane configurations in type IIA string theory, which at low energies realize the linear quivers, are discussed in Section 4 and their M-theory limit is presented in Section 5. We end with some comments and a few problems for the future in Section 6.

\textbf{Note added:} While we were preparing the manuscript the preprint \cite{Xie:2013gma} appeared on the arXiv. There is some overlap between part of our results and the discussion in Section 3 of \cite{Xie:2013gma}.

\section{Linear quivers}
\label{sec:Setup}

\subsection{Setup and symmetries}

\begin{figure}[ht]
 \centering
\includegraphics[scale=1]{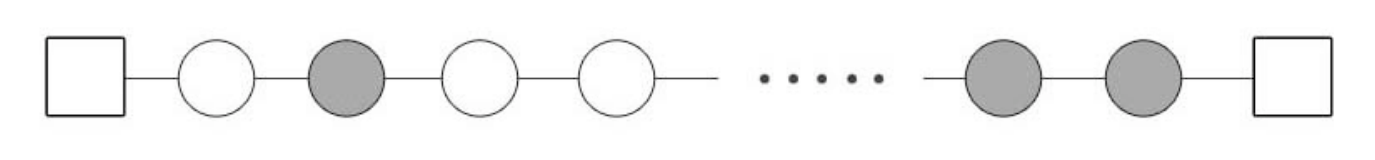}
\caption{\textit{A general linear quiver. The shaded and unshaded circles denote $\mathcal{N}=1$ and $\mathcal{N}=2$ vector multiplets respectively. The lines connecting them are hypermultiplets in the bifundamental of the $SU(N)$ gauge groups at the two ends of the line. The boxes at both ends of the quiver represent two sets of $N$ hypermultiplets in the fundamental of the $SU(N)$ gauge group.}}
\label{LinearQ}
\end{figure}

The aim of this paper is to understand the IR dynamics of linear quivers with gauge group $G$ which is a product of $\ell-1$ copies of $SU(N)$
\begin{equation}
G = \prod_{i=1}^{\ell-1} SU(N)\;.
\end{equation}  
The general quiver we have in mind is illustrated in Figure \ref{LinearQ}.  The matter content of the field theory is encoded in the shaded quiver diagram as follows
\begin{itemize}
\item Shaded circles correspond to $SU(N)$ gauge groups with $\mathcal{N}=1$ vector multiplets.  There are $n_1$ of them.
\item Unshaded circles correspond to $SU(N)$ gauge groups with $\mathcal{N}=2$ vector multiplets, i.e. an $\mathcal{N}=1$ vector multiplets with an adjoint chiral superfield.  There are $n_2$ of them.
\item Lines between circles correspond to $SU(N)\times SU(N)$ bifundamental hypermultiplets.  There are $\ell-2$ of them.
\item The boxes at the end of the quiver diagram correspond to two sets of $N$ hypermultiplets in the fundamental representation of the two end $SU(N)$ gauge groups.  
\end{itemize}  
We have a total of $n_1+n_2=\ell-1$ gauge groups and $\ell$ hypermultiplets.  We use $V_i$ to denote the $i$th gauge group, with $i=1$ corresponding to the left most circle.  Let $\phi_i$ denote the chiral adjoint in the $i$th vector multiplet. If the $i$th vector multiplet is $\mathcal{N}=1$ there is no $\phi_i$ field.  As usual, the hypermultiplets consist of a pair of chiral superfields in conjugate representations, we denote the full hypermultiplet as $H_i=(Q_i,\widetilde{Q}_i)$ with $i=0$ corresponding to the left box and $i=\ell$ corresponding to the right box.  

The quivers of interest possess large global symmetry in addition to the $\mathcal{N}=1$ supersymmetry.  There is an $SU(N)$ flavor symmetry acting on the end hypermultiplets and a $U(1)$ flavor symmetry for each $H_i$ and $\phi_i$.  There is also an overall $R$-symmetry.  The global symmetry is therefore
\begin{equation}
SU(N)\times SU(N) \times U(1)^{\ell + n_2} \times U(1)_{R}\;.  
\end{equation} 
We denote the $U(1)$ symmetries acting on the hypermultiplets as $J_i$ and those acting on the chiral adjoints (when they are present) as $F_i$.  We normalize the charges as
\begin{equation}
J_i (Q_j)=J_i(\widetilde{Q}_j) = \delta_{ij}\;, \qquad F_i(\phi_j) = \delta_{ij}\;.  
\end{equation}
Some of these global $U(1)$ symmetries suffer from chiral anomalies.  Each gauge group yields one anomaly constraint and therefore we can construct $n_2+1$ anomaly free $U(1)$'s.  We also have an anomaly free $R$-symmetry denoted as $R_0$.  We can choose the charge assignments for the $R$-symmetry as
\begin{equation}
R_0 (Q_i) = R_0 (\widetilde{Q}_i) = \frac{1}{2}\;, \qquad R_0(\phi_i) =1\;.  
\end{equation}
%

\subsection{Quivers without superpotential}

Without any superpotential terms, we expect the quiver to break into $n_2+1$ smaller quivers in the IR.  The one-loop beta functions for the gauge group couplings are 
\begin{equation}
b_0(V_i^{{\mathcal{N}=2}})=0\;, \qquad b_0(V_i^{{\mathcal{N}=1}} )= -N\;. \label{betaf}
\end{equation}  
The gauge couplings for the $\mathcal{N}=2$ gauge groups are marginal.  Without any superpotential terms we expect these gauge couplings to be marginally irrelevant \cite{Green:2010da}.  As a result the $\mathcal{N}=2$ gauge groups are non-dynamical and therefore the quiver breaks up at these sites in the IR to yield $n_2+1$ smaller quivers.

\begin{figure}[ht]
 \centering
\subfigure[]{\label{Simp1}\includegraphics[scale=1]{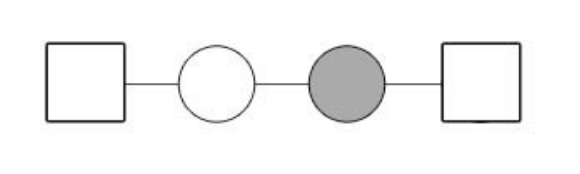}}
\subfigure[]{\label{Simp1Dual}\includegraphics[scale=1]{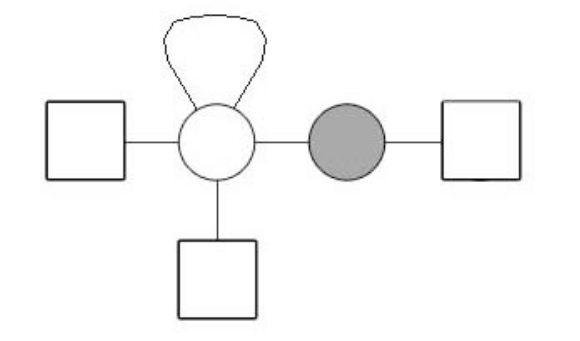}}
\caption{\textit{A simple linear quiver with $\ell=3$ (a) and its Seiberg dual when there is no superpotential turned on (b). After Seiberg duality the mesons charged under the $\mathcal{N}=2$ vector multiplet are represented as a fundamental hypermultiplet and a chiral adjoint.}}
\end{figure}

To illustrate this point, we consider the simple quiver in Figure \ref{Simp1} and try to follow the dynamics as we flow to the IR.  From the one-loop beta functions in (\ref{betaf}), we expect the $\mathcal{N}=1$ vector in Figure \ref{Simp1} to become strongly coupled while the $\mathcal{N}=2$ vector stays weakly coupled.  We can then Seiberg dualize at the $\mathcal{N}=1$ node.  The gauge group will still be $SU(N)$ since $N_f=2N_c$ locally. After the Seiberg duality the mesons of the electric theory become fundamental fields charged under the $\mathcal{N}=2$ gauge group.  These ``mesons'' will decompose into $N$ hypermultiplet in the fundamental representation of the $SU(N)$ $\mathcal{N}=2$ vector multiplet and an adjoint chiral superfield.  The resulting quiver is depicted in Figure \ref{Simp1Dual}.  There is also a superpotential term generated which couples the new fields to the bifundamental hypermultiplets at the $\mathcal{N}=1$ vector multiplet.  The one loop beta function for the $\mathcal{N}=2$ gauge group is then 
\begin{equation}
b_0(V_i^{{\mathcal{N}=2}} ) = 2N\;.
\end{equation}  
The gauge coupling is therefore irrelevant and the $\mathcal{N}=2$ vector multiplet has no interesting dynamics in the IR.  We expect the $\mathcal{N}=2$ vectors adjacent to the $\mathcal{N}=1$ vectors in the general linear quiver in Figure \ref{LinearQ} to behave in the same way. Thus in the absence of superpotential terms the general linear quiver of Figure \ref{LinearQ} will break into $n_2+1$ decoupled smaller quivers. Our discussion fits well with the known fact that $\mathcal{N}=2$ vector multiplets coupled to $N_f=2N_c$ matter run free in the IR when the $\mathcal{N}=2$ superpotential term is not present \cite{Strassler:fk}.

\subsection{Quivers with superpotential}

We are interested in situations where the IR dynamics of the quivers in Figure \ref{LinearQ} is non-trivial. More precisely we will argue that for appropriate choice of the superpotential the physics in the IR is governed by an $\mathcal{N}=1$ superconformal field theory. 

We can avoid the problem of having the quiver break apart by turning on superpotential terms.  At $\mathcal{N}=1$ sites, we turn on 
\begin{equation}
W^i_{\mathcal{N}=1} =\beta_i (Q_{i-1} \widetilde{Q}_{i-1}) (\widetilde{Q}_i Q_i) \label{1spt}\;,
\end{equation} 
where $\beta_i$ are arbitrary complex numbers. At $\mathcal{N}=2$ sites, we can turn the superpotential
\begin{equation}
W^i_{\mathcal{N}=2} = \alpha_{L}^i \phi_i (Q_{i-1} \widetilde{Q}_{i-1}) + \alpha_{R}^i \phi_i (\widetilde{Q}_{i}Q_i) \label{2spt}\;,
\end{equation}
where $\alpha_{L,R}^{i}$ are complex numbers. The superpotential terms in \eqref{1spt} and \eqref{2spt} generate masses for the extra fields introduced after the Seiberg duality depicted in Figure \ref{Simp1Dual}. These superpotential prevents the marginal $\mathcal{N}=2$ gauge coupling from running free. We now study the quiver in Figure \ref{LinearQ} with these superpotential terms.  

The superpotential terms in \eqref{1spt} and \eqref{2spt} preserve the $R_0$ $R$-symmetry, and break all but one of the anomaly free flavor $U(1)$ symmetries of the linear quiver.  In order to find this $U(1)$, we need to understand how the chiral anomaly is cancelled at a given gauge group site.  At the $i$th node of the quiver, the local combination $J_{i-1} -J_{i}$ is always anomaly free.  If the site contains a chiral adjoint then there is an additional anomaly free local $U(1)$ given by $J_{i-1} + J_i -2F_i$.  The superpotential terms at the $\mathcal{N}=2$ sites in \eqref{2spt} break the former local $U(1)$.  The lesson is that the charges of the hypermultiplets flip sign across $\mathcal{N}=1$ vectors and stay the same across $\mathcal{N}=2$ vectors.  We can thus assign to each hypermultiplet a sign $\sigma_i=\pm 1$ and follow the rule that $\mathcal{N}=1$ ($\mathcal{N}=2$) vector multiplets connect hypermultiplets of different (same) sign. The non-anomalous global $U(1)$ symmetry can then be written as
\begin{equation}
\mathcal{F} = \sum_{H_i} \sigma_i J_i - \sum_{V_i} (\sigma_{i-1} + \sigma_i) F_i\;, \label{Fdef}
\end{equation}
where the first sum is over all hypers and the second one is over all vectors. It is straightforward to check that this is the only anomaly free flavor $U(1)$ preserved by the superpotential terms in \eqref{1spt} and \eqref{2spt}.  In general there will be $p$ $J$'s with $\sigma_i=1$ and $q$ $J$'s with $\sigma_i=-1$.  These parameters are constrained to obey $p+q=\ell$. This setup and rules are very similar to the ones used for the generalized $\mathcal{N}=1$ quivers constructed in \cite{Bah:2011je,Bah:2011vv,Bah:2012dg}.

One can also contemplate the addition of the superpotential term of the form
\begin{equation}
W^i_{\mathcal{N}=1} =\gamma_i (\widetilde{Q}_i Q_i)^2 \label{1sptFbreak}\;,
\end{equation} 
for any set of complex numbers $\gamma_i$. This superpotential breaks the $U(1)$ symmetry denoted by $\mathcal{F}$ in \eqref{Fdef}. As we will discuss below when the superpotential \eqref{1sptFbreak} is turned on, the theory always flows to the same IR fixed point. Only when we arrange all the coefficients $\gamma_i$ to vanish we find the extra $U(1)$ global symmetry in \eqref{Fdef} which allows for an interesting family of interacting SCFTs in the IR. We now proceed to study this family of fixed points.

\section{IR dynamics}
\label{sec:IRdynamics}

In this section we will assume that the IR dynamics of the linear quivers with the superpotential terms in (\ref{1spt}) and (\ref{2spt}) is controlled by a superconformal field theory and perform a number of consistency checks of this claim.  Our main calculation tool will be the knowledge of the global symmetries together with a-maximization \cite{Intriligator:2003jj}.  

\subsection{Central charges and $R$-symmetry}  

If the linear quiver flows to an IR SCFT there should be a superconformal $R$-symmetry which we can determine by using a-maximization \cite{Intriligator:2003jj}.  Once we know this $R$-symmetry, we can determine dimensions of chiral operators and check unitarity bounds.  We can also compute the central charges of the theory.  

 If we have a superconformal fix point, the $a$ and $c$ central charges are given by the 't Hooft anomalies associated with the superconformal $R$-symmetry \cite{Anselmi:1997am}, $R_{\mathcal{N}=1}$
\begin{equation}
a = \frac{3}{32} \left(3 \mbox{Tr}R^3_{\mathcal{N}=1} - \mbox{Tr}R_{\mathcal{N}=1} \right)\;, \qquad c = \frac{1}{32} \left(9 \mbox{Tr}R^3_{\mathcal{N}=1}  - 5\mbox{Tr}R_{\mathcal{N}=1}  \right)\;. \label{centralch}
\end{equation} 
The linear quivers admit a one-parameter family of $R$-symmetries which are linear combinations of $R_0$ and $\mathcal{F}$  
\begin{equation}
R_\epsilon = R_0 + \frac{1}{2} \epsilon \mathcal{F}\;.
\end{equation} 
The real number $\epsilon$ is apriori unknown. Each $R_\epsilon$ yields an $a(\epsilon)$ via (\ref{centralch}).  The superconformal R-symmetries maximizes the function $a(\epsilon)$ and thus uniquely determines the value of $\epsilon$ \cite{Intriligator:2003jj}.  Now we proceed with the calculation of the 't Hooft anomalies from the vector and matter multiplets of the linear quiver.

The charges of the superfields are
\begin{equation}
R_\epsilon(Q_i)=R_\epsilon(\widetilde{Q}_i) = \frac{1}{2} (1 + \epsilon \sigma_i)\;, \qquad \mbox{and} \qquad R_\epsilon(\phi_i) = 1-\frac{1}{2} \epsilon (\sigma_{i-1} +\sigma_i)\;.  
\end{equation}
The 't Hooft anomalies are
\begin{align}
\mbox{Tr}R_\epsilon^3(H_i) = \frac{1}{4} N^2 (\epsilon \sigma_i -1)^3\;, \qquad \mbox{Tr}R_\epsilon (H_i)  = N^2 (\epsilon \sigma_i -1)\;,
\end{align}
for the $i$th hypermultiplet and
\begin{equation}
\begin{split}
\mbox{Tr}R_\epsilon^3(V_i) &=  (N^2-1) \left[1-\frac{1}{8} \epsilon^3 (\sigma_{i-1} +\sigma_i)^3\right]\;, \\
 \mbox{Tr}R_\epsilon (V_i)  &= (N^2-1) \left[1-\frac{1}{2} \epsilon (\sigma_{i-1} +\sigma_i)\right]\;,
 \end{split}
\end{equation}
 for the $i$th vector multiplet.  

We can write the total anomaly by summing over all fields in the quiver and obtain
\begin{equation}
\begin{split}
\mbox{Tr}R (H) & = \ell N^2 \left(z \epsilon -1\right) \;,\\
\mbox{Tr}R^3(H) &= \frac{1}{4} \ell N^2 \left( z (3 \epsilon + \epsilon^3) - (1+ 3 \epsilon^2) \right)\;,
\end{split}
\end{equation}
for the hypermultiplets, and
\begin{equation}\begin{split}
\mbox{Tr}R (V) & =  (N^2-1) \left( \ell-1 -\epsilon (z \ell -\kappa) \right) \;,\\
\mbox{Tr}R^3(V) &=  (N^2-1) \left( \ell-1 -\epsilon^3 (z \ell -\kappa) \right) \;,
\end{split}
\end{equation}
for the vector multiplets.  We have defined two new parameters
\begin{equation}
z = \frac{p-q}{\ell}\;, \qquad \mbox{and} \qquad \kappa = \frac{1}{2} (\sigma_0 + \sigma_\ell)\;, \label{zkappa}
\end{equation} 
these parameters are important for labeling different SCFTs.

The trial central charge $a(\epsilon)$ is
\begin{equation}
a(\epsilon) = \frac{3}{4\cdot 32} \left[3 A_3 \epsilon^3 -9 A_2 \epsilon^2 + A_1 \epsilon + A_0\right]\;, \label{aeps}
\end{equation}
where
\begin{equation}
\begin{split}
A_3 &=4 \kappa (N^2-1) + z \ell (4-3N^2)\;, \qquad A_2 = N^2 \ell\;, \\
A_1 &= z \ell (9N^2-4) - 4 \kappa (N^2-1)\;, \qquad A_0 = \ell N^2 + 8(N^2-1)(\ell-1)\;.
\end{split} 
\end{equation}
The function $a(\epsilon)$ is maximized at $\epsilon=\epsilon_m$ with
\begin{equation}
\epsilon_m = \frac{3A_2 - \sqrt{9A_2^2 -A_1 A_3}}{3A_3}\;. \label{soleps}
\end{equation}  
The 't Hooft anomalies at the superconformal fix points are given by
\begin{align}
\mbox{Tr}R_{\mathcal{N}=1} & = \ell N^2 \left(z \epsilon_m -1\right) + (N^2-1) \left( \ell-1 -\epsilon_m (z \ell -\kappa) \right)\;, \label{thooft}\\ 
\mbox{Tr}R^3_{\mathcal{N}=1} &= \frac{1}{4} \ell N^2 \left( z (3 \epsilon_m + \epsilon_m^3) - (1+ 3 \epsilon_m^2) \right) +  (N^2-1) \left( \ell-1 -\epsilon_m^3 (z \ell -\kappa) \right)\;,  \nonumber
\end{align} 
where $\epsilon_m$ is given in (\ref{soleps}).  The $a$ and $c$ central charges can be easily deduced from the expressions in (\ref{centralch}).

Each theory in the IR is labelled by the discrete parameters $\{\kappa,z, \ell, N\}$ and the central charges depend only on these parameters. It is natural to conjecture that all linear quivers with the same values of the parameters $\{\kappa,z, \ell, N\}$ are dual to each other and flow to the same IR SCFT.

The parameter $\epsilon_m$ is odd under $(z,\kappa) \to (-z,-\kappa)$, therefore the 't Hooft anomalies and the central charges are invariant under such transformation.  From the definitions of $z$ and $\kappa$ in (\ref{zkappa}) we observe that there are three choices for $\kappa$, $\{-1,0,1\}$, and $|z|$ is bounded above, $|z|\leq 1$.  Without lost generality, we can restrict $z$ to the range $0\leq z \leq 1$.  This correspond to restricting the parameters $p$ and $q$ to obey $q\leq p$.

\subsection{Consistency checks}

\subsubsection*{Unitarity bound}
A consistency check for the validity of $a$-maximization and for the claim that there is an IR SCFT is to make sure that chiral operators satisfy the unitarity bound, i.e.
\begin{equation}
\Delta = \frac{3}{2} R_{\mathcal{N}=1} \geq 1\;.
\end{equation} 
The charges for the fundamental fields in terms of the trial R-symmetry are
\begin{equation}
R_{\mathcal{N}=1}(Q_i) = \frac{1}{2} (1 + \epsilon_m \sigma_i)\;, \qquad \mbox{and} \qquad R_{\mathcal{N}=1}(\phi_i) = 1-\frac{1}{2} \epsilon_m (\sigma_{i-1} +\sigma_i)\;.  
\end{equation}  
The lowest dimensional gauge invariant operators that can be constructed from these are mesons from the hypermultiplets and mass terms for the chiral adjoints.  Their charges are
\begin{equation}
R_{\mathcal{N}=1}(Q_i \widetilde{Q}_i) = 1+ \epsilon_m \sigma_i \;,\qquad  R_{\mathcal{N}=1}(\phi_i^2) = 2- \epsilon_m (\sigma_{i-1} + \sigma_{i})\;.
\end{equation} 
The unitarity bound is obeyed when
\begin{equation}
-\frac{1}{3} \leq \epsilon_m \leq \frac{1}{3}\;. \label{epsbounds}
\end{equation}  
One can check that $\epsilon_m$ in \eqref{soleps} always lies within this range for the allowed ranges of the parameters $\{\kappa,z,\ell,N\}$.  The bounds in \eqref{epsbounds} are saturated by $z=-1$ (lower) and $z=1$ (higher).  

\subsubsection*{Hofman-Maldacena bound}

We can also check a number of non-trivial bounds obeyed by the central charges of any $\mathcal{N}=1$ SCFT. It is not hard to show that for all allowed values of the parameters $\{\kappa,z,\ell,N\}$ both $a$ and $c$ are positive. One can also show that the Hofman-Maldacena bound for $\mathcal{N}=1$ SCFT's \cite{Hofman:2008ar} is obeyed, i.e.
\begin{equation}
\frac{1}{2} \leq \frac{a}{c} \leq \frac{3}{2}\;. \label{HMbound}
\end{equation}
In fact for the linear quivers studied here we find a narrower range
\begin{equation}
\frac{1}{2} \leq \frac{a}{c} \leq 1\;.  \label{acbound}
\end{equation}
The lower bound is obtained by setting $\ell =1$.  For this case, there are no vector multiplets and one has $z=1$.  The central charges for $\ell=1$ are 
\begin{equation}
c=2a = \frac{1}{24} \left[(3-\kappa) N^2 -(1-\kappa) \right]. \label{achalf}
\end{equation} 
When $\kappa=1$ the central charge is just that of an $SU(N)\times SU(N)$ bifundamental hypermultiplet.  This is not surprising since $\ell=z=1$ for $\kappa=1$ is precisely the theory of a 4D $SU(N)\times SU(N)$ bifundamental hypermultiplet.  It is also consistent with the fact that the lower limit of the Hofman-Maldacena bound \eqref{HMbound} is saturated by free hypermultiplets.  

The theories with $z=\ell=1$ and $\kappa =0,-1$ (for any positive $N$) are more mysterious. We cannot construct these theories as ordinary linear quivers of the type discussed in Section \ref{sec:Setup}. However the central charges for these values of the parameters obey all constraints for describing a good $\mathcal{N}=1$ SCFT and we will see in Section \ref{sec:M-theory} that there is also a nice M-theory picture which suggests that the theories with $z=\ell=1$ and $\kappa =0,-1$ should be taken seriously as new SCFTs without any known 4D Lagrangian description. In fact we believe that this phenomenon may be more general, i.e. there are values of the parameters $\{\kappa,z,\ell,N\}$ for which the linear quiver description does not make sense but if the M-theory construction of Section \ref{sec:M-theory} is sensible and the central charges obey all consistency conditions we should probably view these theories as good $\mathcal{N}=1$ SCFTs.

The upper bound in \eqref{acbound} is saturated in the large $\ell$ limit. The theories in this limit may admit holographic duals.  It is amusing that none of our SCFTs have $a>c$.  This may not be too surprising after recalling that upper limit of the Hofman-Maldacena bound \eqref{HMbound} is saturated by free vector multiplets and in the linear quivers we cannot isolate a limit in which the effective degrees of freedoms are only vectors.  

\subsubsection*{Large $N$ limit}

In the large $N$ limit for $\kappa=0$ one finds
\begin{equation}
\begin{split}
a&=\dfrac{N^2}{64}\dfrac{((1+3z^2)^{3/2}+9z^2-1)\ell-12z^2}{z^2}\;, \\
c&=\dfrac{N^2}{64}\dfrac{((1+3z^2)^{3/2}+9z^2-1)\ell-8z^2}{z^2}\;.
\end{split}
\end{equation}
For $\kappa=\pm1$ in the large $N$ limit the expressions for the central charges are unwieldy but one again finds that for finite $\ell$ one has $a\neq c$. It is interesting that in the large $N$ \textit{and} large $\ell$ limit (keeping $z$ fixed) one finds
\begin{equation}
a=c=\dfrac{\ell N^2}{64}\dfrac{((1+3z^2)^{3/2}+9z^2-1)}{z^2} \;.
\end{equation}
Note that the dependence on $z$ in this limit is precisely the same as the one found in the large $N$ limit for the SCFTs studied in \cite{Bah:2011vv,Bah:2012dg}. The fact that we have $a=c$ in this limit also suggests that these SCFTs may admit a holographic dual description in type IIA or 11D supergravity. Curiously for $\ell=4(g-1)N/3$ we get exactly the same numerical values of the central charges as for large $N$ limit of the theories in \cite{Bah:2011vv,Bah:2012dg} coming from hyperbolic Riemann surfaces.

\subsubsection*{Universal RG flow}

It was shown in \cite{Tachikawa:2009tt} that if a UV SCFT with $\mathcal{N}=2$ supersymmetry is deformed by a mass term for the chiral adjoint in the $\mathcal{N}=2$ vector multiplet and the theory flows to an $\mathcal{N}=1$ SCFT in the IR then there is a universal relation between the central charges in the IR and UV given by
\begin{equation}
a_{\text{IR}} = \dfrac{9}{32}(4a_{\text{UV}}-c_{\text{UV}})\;, \qquad\qquad c_{\text{IR}} = \dfrac{1}{32}(-12a_{\text{UV}}+39c_{\text{UV}}) \;. \label{BWidentities}
\end{equation}
One can show that these identities are obeyed if the UV theory is the one with $z=1$ and $\kappa=1$ and the IR one is the one with $z=0$ and $\kappa=0$. These theories are precisely the two theories for which one does not need $a$-maximization as a result of which the central charges are rational and it is natural to conjecture that they are related by the universal RG flow of \cite{Tachikawa:2009tt}. The exact expressions for the central charges are
\begin{equation}
\begin{split}
a_{z=0,\kappa=0} &= \dfrac{3}{128}[N^2(9\ell-8)-8(\ell-1)]\;, ~~~ c_{z=0,\kappa=0} = \dfrac{1}{128}[N^2(27\ell-16)-16(\ell-1)]\;,\\
a_{z=1,\kappa=1} &= \dfrac{1}{24}[N^2(6\ell-5)-5(\ell-1)]\;, ~~~ c_{z=1,\kappa=1} = \dfrac{1}{12}[N^2(3\ell-2)-2(\ell-1)]\;, \notag
\end{split}
\end{equation}
and it is easy to check that they obey \eqref{BWidentities}.

\subsection{Dualities and conformal manifold}\label{conm}

The SCFTs obtained from the linear quivers are labelled by four parameters, $\{\kappa,z, \ell, N\}$. For a given SCFT, we can find more than one way to construct the UV linear quiver by changing the relative number of $\mathcal{N}=2$ and $\mathcal{N}=1$ vector multiplets. We, therefore, observe an interesting version of Seiberg duality for these SCFTs. A similar duality was observe in the field theory constructions of the SCFTs in the IR of M5-branes on Riemann surface \cite{Bah:2011vv,Bah:2012dg,Gadde:2013fma}. Unlike the M5-brane constructions, we have explicit Lagrangian description for the linear quiver theories and thus one can study and understand these dualities in greater detail. We leave this detailed analysis for the future.

We can compute the dimension of the conformal manifold for the IR SCFTs using the method of Leigh-Strassler \cite{Leigh:1995ep} (see \cite{Green:2010da} for a modern incarnation of this method). There are $\ell-1$ complex gauge couplings, $n_1$ complex superpotential couplings from the $\mathcal{N}=1$ vectors and $2n_2$ complex couplings from the $\mathcal{N}=2$ vectors. The number of constraints are given by the number of anomalous $U(1)$'s which is $\ell + n_2$. This yields a total of $\ell-1$ exactly marginal complex parameters. If we allow the superpotential terms associated to the box hypermultiplets in the linear quiver that break the global $SU(N)\times SU(N)$ symmetry the conformal manifold would be even larger since then one finds $2(N^2-1)$ additional exactly marginal parameters.

\section{Type IIA construction}  
\label{sec:IIA}

\begin{figure}[ht]
 \centering
\includegraphics[scale=.8]{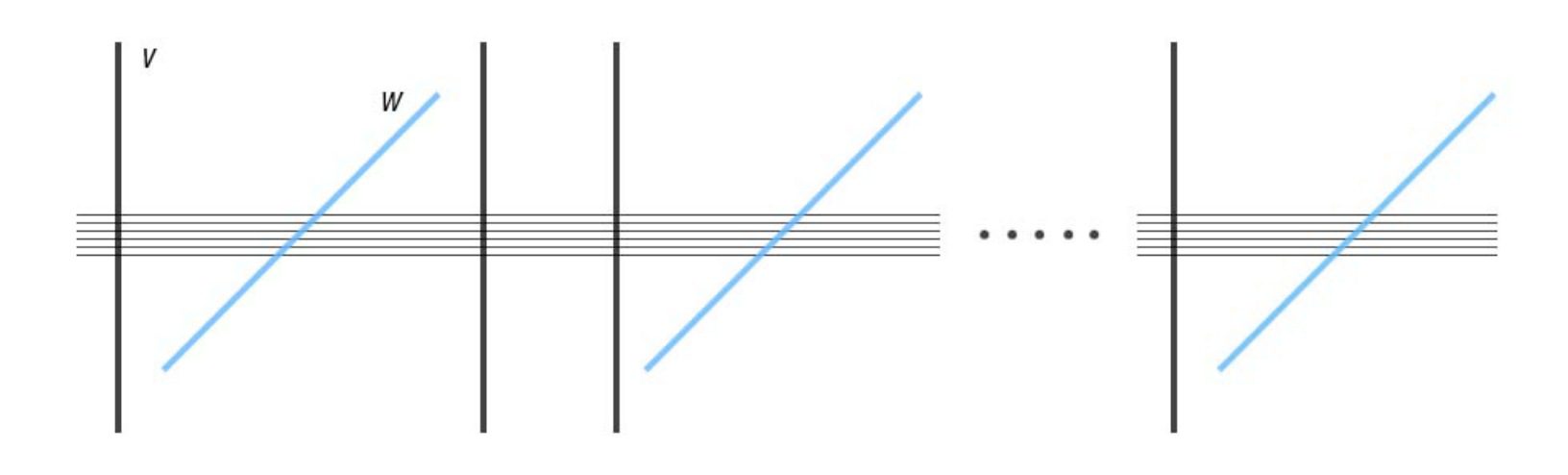}
\caption{\textit{A configuration of intersecting D4- and NS5-branes which corresponds to a linear quiver. The horizontal black lines represent a stack of $N$ D4-branes extended along the $x_6$ direction. The vertical black lines are the $v$ NS5-branes extended along $x_{4,5}$. The blue lines represent the $w$ NS5-branes extended along $x_{7,8}$.  All branes extend along the 4D space-time directions $x_{0,1,2,3}$ and are localized at $x_9=0$. }}
\label{LinearBranes}
\end{figure}

The linear quivers, above, can be obtained in type IIA string theory as the low energy and weak coupling limit of intersecting D4- and NS5-branes. This construction is very similar to the $\mathcal{N}=2$ linear quivers studied in \cite{Witten:1997sc}.  We take the ten space-time coordinates to be $x_{0,1,\cdots, 9}$, with $x_0$ being time.  We consider $N$ coincident D4-branes extended along $x_{0,1,2,3,6}$ and sitting at the point $x_{4,5,7,8,9}=0$.  We add $p$ non-coincident NS5-branes extended along $x_{0,1,2,3,4,5}$ and localized at $x_{7,8,9}=0$. Each of these branes is also localized at a point$x_6=x_6^{\alpha}$, where $\alpha$ is an integer in the set $\{1,\ldots,p\}$.  We also add $q$ non-coincident NS5-branes extended along $x_{0,1,2,3,7,8}$, localized at $x_{4,5,9}=0$ and each of them sitting at a point $x_6=x_6^{\beta}$, where $\beta$ is an interger in $\{1,\ldots,q\}$.  The total number of NS5-branes is then $\ell=p+q$.  We do not assume any particular ordering of the NS5-branes along the $x_6$ direction.  We illustrate an example of a brane configuration of this type in Figure \ref{LinearBranes}.   

We introduce the complex coordinates $v=x_4+ix_5$ and $w=x_7+ix_8$ and call the NS5-branes extended along $x_{4,5}$, $v$-branes, and those along $x_{7,8}$, $w$ branes.  Between any two adjacent NS5-branes, there is a suspended stack of $N$ D4-branes.  At long distances and weak coupling, there is a four-dimensional $SU(N)$ gauge theory, living on the non-compact part of the D4-brane worldvolume, $x_{0,1,2,3}$, describing the dynamics.\footnote{The gauge theory on the worldvolume of $N$ coincident D4-branes is $U(N)$. As discussed in detail in \cite{Witten:1997sc} due to the presence of the NS5-branes a $U(1)$ subgroup decouples and one is left with an $SU(N)$ gauge group.}  If the NS5-branes are parallel (both are $v$-branes or $w$-branes) there is an additional $SU(N)$ chiral adjoint superfield at low energies corresponding to the freedom of sliding the D4's along the NS5-branes in the $v$ or $w$ directions.  Thus between two parallel NS5-branes we obtain a full $\mathcal{N}=2$ $SU(N)$ vector multiplet.  If the two adjacent NS5-branes are perpendicular (one is a $v$-brane and the other is a $w$-brane) we cannot slide the D4's without a cost in energy, therefore there is only a $\mathcal{N}=1$ $SU(N)$ vector multiplet describing the dynamics at low energies.  At a given NS5 site, there are strings between adjacent D4-branes.  At low energies and weak coupling, they are described by bifundamental hypermultiplets.  Finally, there are two semi-infinite stacks of D4-branes connected to the NS5-branes at the two ends of the brane system.  The gauge groups associated to these D4-branes are frozen and thus we are left with two sets of $N$ hypermultiplets in the fundamental representation of $SU(N)$ coming from the strings at the NS5-branes at the two ends. It is now clear that this collection of intersecting branes realizes the gauge fields and matter content of the linear quivers described in Section \ref{sec:Setup}.

The map between the field theory and brane constructions can be made more precise.  It is clear that to each bifundamental hypermultiplet there is an associate NS5-brane.  The hypermultiplets with $\sigma_i=1$ in Section \ref{sec:Setup} can be associated with the $v$-branes, and the hypermultiplets with $\sigma_i=-1$ can be associated with the $w$-branes.  There is a $U(1)_v$ and a $U(1)_w$ symmetry acting on the $v$ and $w$ plane respectively.  These $U(1)$'s manifest themselves in the quiver as local $U(1)$ $R$-symmetry acting on the hypermultiplets and chiral adjoints. In terms of the symmetries defined in Section \ref{sec:Setup} we have
\begin{equation}
R_0 = U(1)_v+U(1)_w\;, \qquad \mathcal{F} =  U(1)_v-U(1)_w\;.
\end{equation}

In the weak coupling limit, the inverse gauge coupling of the gauge field between two adjacent NS5-branes is proportional to the distance between them \cite{Witten:1997sc}.  Since we are free to pick the positions of the NS5-branes, the distances between the branes are marginal parameters.  At strong coupling, we cannot describe the gauge couplings in this way since the branes recombine at the intersections.  However, as we move far way from the intersection region, superconformal symmetry imposes the condition that the NS5-branes should not bend \cite{Witten:1997sc}. Thus the asymptotic behaviour of the NS5-branes must stay the same as in the weak coupling limit.  The distances between the branes, far away from the intersection region, must correspond to exactly marginal parameters. If we have $\ell$ NS5-branes then there are $\ell-1$ distances we can freely choose and thus exactly $\ell-1$ marginal parameters. These parameters are real but as discussed in \cite{Witten:1997sc} and in the next section when we take the M-theory limit the distance along the M-theory circle $x_{10}$ naturally complexifies the $x_6$ distance and leads to $\ell-1$ complex marginal parameters.   This coincides with the counting of marginal couplings in the field theory discussed in Section \ref{conm}. The brane picture makes it also clear that the IR theory is insensitive to the particular ordering of NS5-branes of type $v$ and $w$ as long as their number is kept fixed. This is one more manifestation of the fact that the IR SCFTs are labelled only by the parameters $\{\kappa,z,\ell,N\}$ and different UV constructions with the same values of these parameters should result in dual descriptions of the same theory.

In the past, there have been many constructions of $\mathcal{N}=1$ field theories that use intersecting D4- and NS5-branes, see for example \cite{Witten:1997fk,Hori:1997ab,deBoer:1997zy,deBoer:1998by,deBoer:1998rm,Giveon:1997sn,Giveon:1998sr,Elitzur:1998ju}.  In all of the these constructions one starts with some brane configuration involving parallel NS5-branes which preserves $\mathcal{N}=2$ supersymmetry and break this to $\mathcal{N}=1$ by rotating the adjacent parallel NS5-branes at some angle.  From the field theory point of view, this rotation corresponds to giving mass to the adjoint chiral superfield in some $\mathcal{N}=2$ vector multiplet.  Integrating out these chiral adjoints generates superpotential terms of the type (\ref{1sptFbreak}) for hypermultiplets.  In the constructions here, we explicitly turn off these superpotential terms by choosing the NS5-branes which are not parallel to be orthogonal to each other. This choice preserves the additional $U(1)$ flavor symmetry \eqref{Fdef} which in turn is responsible for the rich IR dynamics and the family of SCFT's arising from the linear quivers.  

As discussed in some detail in \cite{Witten:1997sc} in the strong coupling limit we can describe the system of intersecting branes in M-theory. We discuss this next.

\section{Uplift to M-theory}  
\label{sec:M-theory}

In the M-theory limit the space-time becomes eleven-dimensional and the extra coordinate $x_{10}$ is in the shape of a circle.  Both the D4- and NS5-branes in the IIA construction, uplift to M5-branes in M-theory.  The NS5-branes become  M5-branes localized on the $x_{10}$ circle while the D4-branes are obtained by compactifying M5-branes on the circle.  As emphasized in \cite{Witten:1997sc} the $x_6$ direction naturally combines with the $x_{10}$ direction into a complex coordinate
\begin{equation}
s= \frac{x_6+i x_{10}}{R} \; ,\qquad \mbox{or} \qquad t= \exp(-s)\;,
\end{equation}
where $R$ is the radius of the M-theory circle.  

In the M-theory uplift of our type IIA brane construction, the D4-branes branes become M5-branes wrapped on an infinite cylinder (or sphere with two punctures) with complex coordinate $t$.  The NS5-branes become M5-branes which intersect this cylinder at points.  The surface wrapped by the M5 branes is a holomorphic curve in $\mathbb{C}^3$.  After a conformal transformation, we can view this curve as a punctured sphere embedded in $\mathbb{C}^3$.  The two ends of the cylinder (or sphere with two punctures) are two maximal punctures (we use the language of Gaiotto \cite{Gaiotto:2009we}) that corresponds to the intersection with two sets of $N$ M5 branes.  There are also simple punctures on the sphere corresponding to uplifted NS5-branes which intersect the sphere at $\ell$ points.  The brane  system in M-theory thus becomes a set of $N$ coincident M5-branes wrapping a two-sphere with two maximal punctures and $\ell$ simple punctures.  The normal bundle to the two sphere is not the cotangent bundle as in \cite{Witten:1997sc,Gaiotto:2009we} but corresponds to a more general embedding in $\mathbb{C}^3$ as discussed in \cite{Bah:2011vv,Bah:2012dg}. This more general normal bundle results in breaking of $\mathcal{N}=2$ supersymetry to $\mathcal{N}=1$.

\begin{figure}[ht]
 \centering
\includegraphics[scale=.8]{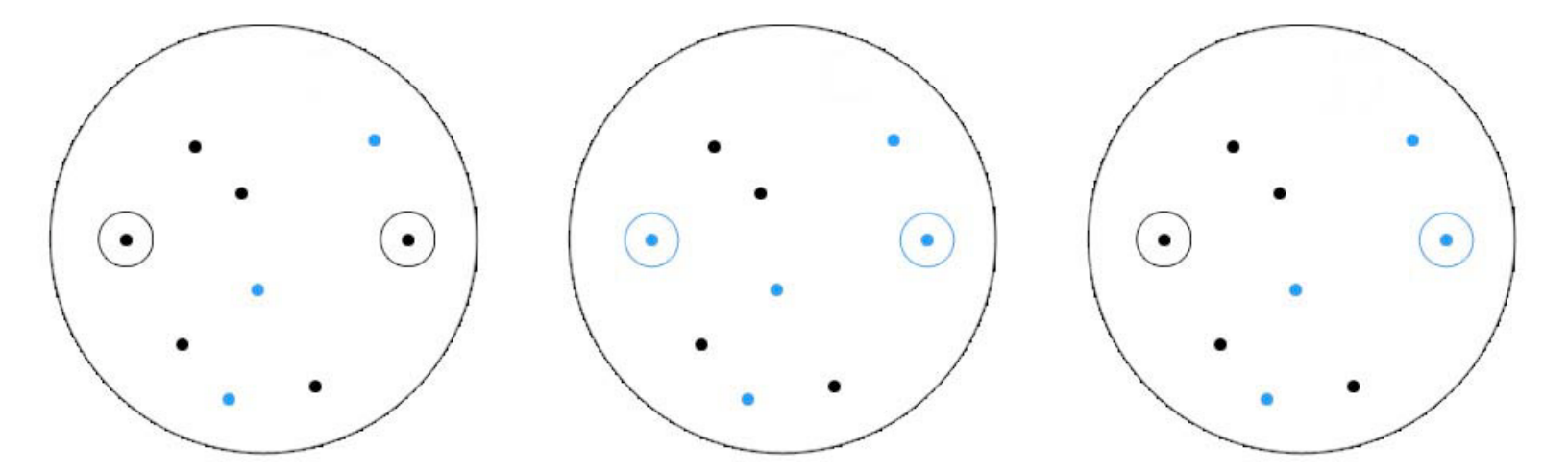}
\caption{\textit{The sphere with seven minimal and two maximal punctures which corresponds to $\ell=7$. We have taken four minimal punctures of type $v$ or $\sigma=+1$ (in black) and three minimal punctures of type $w$ or $\sigma=-1$ (in blue). We have $\kappa=1$ (two black maximal punctures) on the left, $\kappa=-1$ (two blue maximal punctures) in the middle and $\kappa=0$ (one black and one blue maximal puncture) on the right.}}
\label{PunctureSphere}
\end{figure}

An important ingredient in our construction is the presence of two species of punctures corresponding to the fact that some of the punctures come from the uplift of $v$ NS5-branes and some come from $w$ NS5-branes. In the field theory description, this choice is parametrized by the parameter $\sigma_i =\pm1$ which we can now assign to each puncture. We choose to denote the punctures corresponding to $\sigma=1$ with a black dot and the ones with $\sigma=-1$ with a blue dot, see Figure \ref{PunctureSphere}. We have $p$ black dots and $q$ blue ones for a total of $p+q=\ell$ minimal punctures.  In the field theory there is an additional parameter $\kappa$ which labels different theories.  This parameter encodes information about the maximal punctures (denoted by dots with a circle in Figure \ref{PunctureSphere}) which also come in two species (again labeled by blue and black in Figure \ref{PunctureSphere}).  When $\kappa =\pm 1$, the maximal punctures are of the same kind (either blue or black).  When $\kappa=0$, they are of different kind (one blue and one black).  In Figure \ref{PunctureSphere} we illustrate a particular example of a punctured sphere for different values of $\kappa$.

The parameters which determine the IR SCFTs uniquely are $\{\kappa,z, \ell, N\}$, for example only these parameters enter in the central charge.  This fact has the nice geometric interpretation that the relative positions of the punctures on the sphere do not change the IR theory and should correspond to exactly marginal parameters in the SCFT.  We are free to move the punctures around as long as we do not collide them.  We recover different weak coupling limits when we move the punctures far away from each other.  These different limits corresponds to the quivers in the field theory that have different number of $\mathcal{N}=2$ and $\mathcal{N}=1$ vector multiplets for fixed values of $\{\kappa,z, \ell, N\}$. This geometric picture suggests that all linear quivers with the same $\{\kappa,z, \ell, N\}$ are dual to each other and flow to the same SCFT in the IR.

\begin{figure}[ht]
 \centering
\includegraphics[scale=.8]{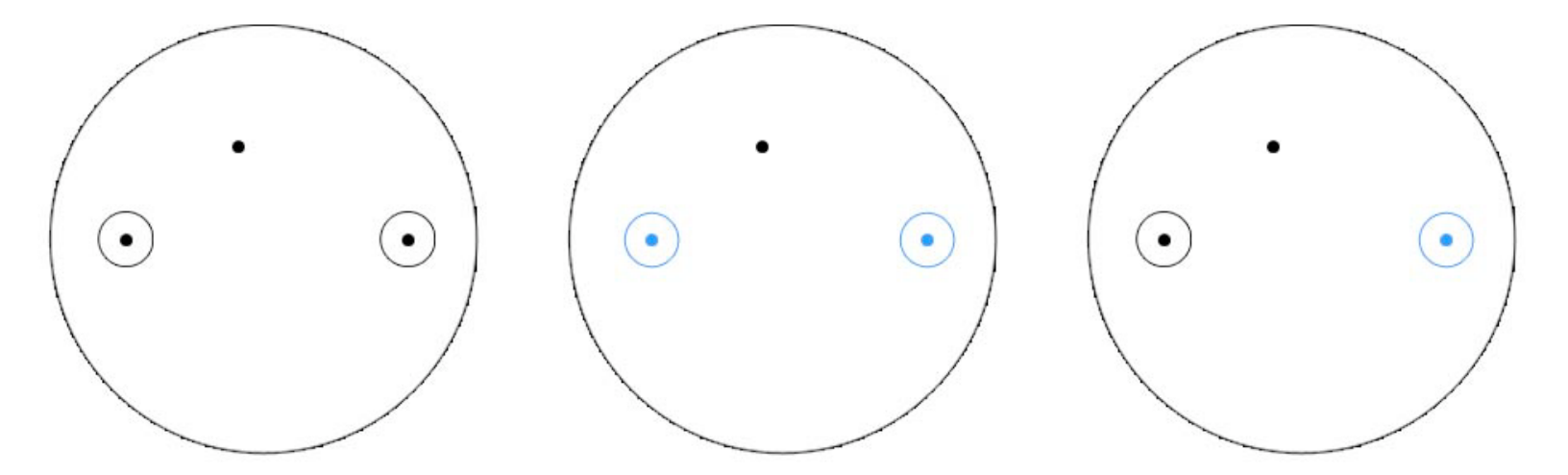}
\caption{\textit{The sphere with two maximal and one minimal puncture. The picture on the left corresponds to a hypermultiplet in the bifundamental of $SU(N)\times SU(N)$. The other two pictures do not have a simple interpretation in field theory but should correspond to isolated $\mathcal{N}=1$ SCFTs.}}
\label{PunctureSphere1}
\end{figure}

The M5-brane picture also suggests the existence of some new SCFTs which can be used as building blocks for constructing more general $\mathcal{N}=1$ quiver theories. To illustrate this point let us consider the punctured sphere with two maximal and one minimal puncture, see Figure \ref{PunctureSphere1}. Without loss of generality we can choose the minimal puncture to correspond to $\sigma=1$, i.e. it is a black puncture. This corresponds to $\ell=z=1$. However we also have the choice of the parameter $\kappa$ (the rank of the gauge group is held fixed). For $\kappa=1$, i.e. two black maximal punctures we can understand the setup in field theory as a hypermultiplet in the bifundamental of $SU(N)\times SU(N)$ and this corresponds to the simplest linear quiver of Section \ref{sec:Setup}. For the other two choices of $\kappa$ we do not have an obvious realization of the field theory in terms of any linear quiver. However it is natural to propose that even for $\kappa=0$ (a blue and a black maximal puncture) and $\kappa=-1$ (two blue maximal punctures) the M5-brane wrapped on this punctured sphere leads to a non-trivial $\mathcal{N}=1$ SCFT. In fact, as discussed around equation (\ref{achalf}), the expressions for the central charges are well-defined and obey all bounds for $\ell=z=1$ and any choice of $\kappa$. Despite the fact that the SCFTs corresponding to $\kappa=0,-1$ and $\ell=z=1$ do not have a known 4D description we can use them as buidling blocks for generalized $\mathcal{N}=1$ quivers that go beyond the linear quivers of Section \ref{sec:Setup} and the IIA brane construction of Section \ref{sec:IIA}. The fact that we know their anomalies, central charges and global symmetry will allow us to get a calculational handle on such generalized quivers even in the absence of an explicit Lagrangian description.

The theories when $\kappa=z=\pm 1$ and any positive $\ell$ and $N$ correspond to having all punctures (minimal and maximal) of the same type.  These theories preserve $\mathcal{N}=2$ supersymmetry.  They admit additional weak couplings limits where the two maximal punctures are brought close to each and this is one way of defining and extracting field theoretic properties of the $\mathcal{N}=2$ $T_N$ theory \cite{Gaiotto:2009we,Gaiotto:2009gz}. It is natural to wonder whether there is a way of defining $\mathcal{N}=1$ generalizations of the $T_N$ theory which can be isolated in a similar way by decoupling some vectors and hypers in the quivers with general values of $\{\kappa,z, \ell, N\}$. We will not offer any specific procedure to achieve this here but would like to point out that one way to study this might be to take limits where the two type of punctures introduced here collide in a controlled way.

\section{Conclusions}

We have argued that a large class of linear-shaped quiver gauge theories with $\mathcal{N}=1$ supersymmetry build out of $\mathcal{N}=1$ and $\mathcal{N}=2$ vector multiplets as well as hypermultiplets have interesting IR dynamics controlled by interacting $\mathcal{N}=1$ SCFTs. We calculated the central charges of these SCFTs and provided some evidence that the linear quivers enjoy a rich set of dualities. These dualities as well as other properties of the field theories are encoded in a brane construction in type IIA string theory or M-theory.

There are clearly many interesting questions for further study. Here we list a few of them.

In this paper we restricted our attention to quivers with linear shape. As pointed out in \cite{Witten:1997sc} for quivers with $\mathcal{N}=2$ supersymmetry the field theory dynamics is modified when one introduces a gauge group that gauges the two fundamental hypermultiplets denoted by boxes in the quiver diagrams in Section \ref{sec:Setup}. This gauging results in a quiver with a circular shape and it will be very interesting to perform a detailed study of such circular quivers with $\mathcal{N}=1$ supersymmetry. It is natural to expect that these will flow to new $\mathcal{N}=1$ SCFTs in the IR. In M-theory the circular quivers should correspond to M5-branes wrapped on a punctured torus.

It should be possible to calculate explicitly the superconformal index of \cite{Romelsberger:2005eg,Kinney:2005ej} for the linear quivers studied here. It should also be possible to uncover some TQFT structure, similar to the one studied in \cite{Gadde:2009kb,Beem:2012yn}, underlying the superconformal index of these theories.

The geometric construction of the linear quivers discussed here in terms of M5-branes wrapped on punctured Riemann surface paves the way for addressing a number of interesting questions.  In the case of $\mathcal{N}=2$ theories, explicit knowledge of the curve wrapped by the M5-branes allowed for a derivation of the Seiberg-Witten curve of the $\mathcal{N}=2$ theory from M-theory \cite{Witten:1997sc}.  Recently this M5-brane construction and the curve wrapped by the M5-branes was instrumental in the pioneering work of \cite{Gaiotto:2009hg,Gaiotto:2009we} which lead to new understanding of the space of $\mathcal{N}=2$ theories and their properties. When we have only $\mathcal{N}=1$ supersymmetry knowledge of the curve wrapped by the M5-branes can still be useful. For example it allowed for the description of the moduli space of SQCD in \cite{Witten:1997fk,Hori:1997ab}.  Even non-holomorphic data can be extracted from this curve as was done in \cite{deBoer:1997zy,deBoer:1998by}. A natural question arising from our construction is thus to understand in more detail the physical information encoded in the punctured sphere wrapped by the M5-branes which leads to our linear quivers.  This may lead to a nice geometric derivation of the $\mathcal{N}=1$ curve of \cite{Intriligator:1994sm} associated with the linear quivers.  Moreover, the punctured sphere and the M5-brane picture may provide us with non-holomorphic data, such as the K\"ahler potential.  By considering various degeneration limits of the punctured surface wrapped by the M5-branes one can explore a larger space of isolated $\mathcal{N}=1$ SCFTs as done for $\mathcal{N}=2$ theories in \cite{Gaiotto:2009we}.  Some questions regarding the $\mathcal{N}=1$ curve of \cite{Intriligator:1994sm} for some generalized $\mathcal{N}=1$ quivers including the $T_N$ theories were studied recently in \cite{Maruyoshi:2013hja}.  

As we discussed in Section \ref{sec:IRdynamics} in the large $\ell$ limit the $a$ and $c$ central charges of the IR SCFTs are equal. This suggests that these theories may admit a holographic dual description in type IIA or 11D supergravity. Gravity duals of $\mathcal{N}=1$ SCFTs arising from M5 branes have been studied before \cite{Maldacena:2000mw,Gaiotto:2009gz,Bah:2011vv,Bah:2012dg} and the underlying brane construction played an instrumental role in the construction of these solutions. It is very likely that the techniques for constructing $AdS_5$ $\mathcal{N}=1$ solutions of M-theory introduced in \cite{Gauntlett:2004zh} and exploited recently in \cite{Bah:2013qya}, will be useful in finding these supergravity solutions.

It will be very interesting if we can isolate a new $\mathcal{N}=1$ building block akin to the $T_N$ theory by going to some special region in the conformal manifolds of our linear quiver theories. This new theory will be interesting in its own right and may provide a new building block for constructing $\mathcal{N}=1$ generalized quiver theories in the spirit of Gaiotto \cite{Gaiotto:2009we}. The $\mathcal{N}=1$ analog of the $T_N$ theory may also provide the missing ingredient for the construction of the SCFTs duals to the infinite set of $AdS_5$ solutions of M-theory found in \cite{Bah:2011vv,Bah:2012dg}.


\acknowledgments

We would like to thank Chris Beem, Francesco Benini, Ken Intriligator, Jaewon Song and Brian Wecht  for many useful discussions during the gestation stage of this project. We acknowledge the warm hospitality provided by the Centro de Ciencias de Benasque Pedro Pascual during the preparation of the manuscript. IB is grateful for the hospitality and work space provided by the UCSD Physics Department and would like to thank Nick Halmagyi for vital logistic support. NB would like to thank his family for crucial support in the final stages of the preparation of the manuscript. IB is supported in part by ANR grant 08-JCJC-0001- 0 and the ERC Starting Grants 240210 - String-QCD-BH, and 259133 - ObservableString. The work of NB was supported in part by the DOE grant DE-FG02-92ER-40697.

\bibliographystyle{utphys}
\bibliography{miracle}

\end{document}